\documentclass[twocolumn,aps,prd,showpacs,amsmath,amssymb,superscriptaddress,nofootinbib]{revtex4}
\usepackage[]{latexsym}
\usepackage[english]{babel}
\usepackage{graphicx}
\usepackage{hyperref}
\usepackage{color}
\renewcommand{\d}{\mathrm{d}}

\begin{document}

\title{Transition redshift in $f(T)$ cosmology and observational constraints}

\author{Salvatore Capozziello}
\affiliation{Dipartimento di Fisica, Universit\`a di Napoli ''Federico II'', Via Cinthia 9,
I-80126 Napoli, Italy.}
\affiliation{Istituto Nazionale di Fisica Nucleare (INFN), Sez. di Napoli, Via Cinthia 9,
I-80126 Napoli, Italy.}
\affiliation{Gran Sasso Science Institute (INFN), Viale F. Crispi 7,  I- 67100 L'Aquila, Italy.}

\author{Orlando Luongo}
\affiliation{Dipartimento di Fisica, Universit\`a di Napoli ''Federico II'', Via Cinthia 9,
I-80126 Napoli, Italy.}
\affiliation{Istituto Nazionale di Fisica Nucleare (INFN), Sez. di Napoli, Via Cinthia 9,
I-80126 Napoli, Italy.}
\affiliation{Instituto de Ciencias Nucleares, Universidad Nacional Autonoma de M\'exico
(UNAM), Mexico.}

\author{Emmanuel N.  Saridakis}
\affiliation{Physics Division, National Technical University of Athens, Zografou Campus,
Athens, Greece.}
\affiliation{Instituto de F\'{\i}sica, Pontificia, Universidad de Cat\'olica de
Valpara\'{\i}so, Casilla, Valpara\'{\i}so, Chile.}

\begin{abstract}
We extract constraints on the transition redshift $z_{tr}$, determining the onset of
cosmic acceleration, predicted by an effective cosmographic construction, in the framework
of $f(T)$ gravity. In particular, employing cosmography we obtain bounds on the viable
$f(T)$ forms and their derivatives. Since this procedure is model independent, as long as
the scalar curvature is fixed, we are able to determine intervals for $z_{tr}$. In this
way we guarantee that the Solar-System constraints are preserved and moreover we extract
bounds on the transition time and the free parameters of the scenario.
%We phenomenologically reconstruct an effective Hubble rate derived from numerically
%solving the modified Friedmann equations by means of cosmographic initial settings.
We find that the transition redshifts predicted by $f(T)$ cosmology, although compatible
with the standard $\Lambda$CDM predictions, are slightly smaller. Finally, in order to
obtain observational constraints on $f(T)$ cosmology, we perform a Monte Carlo fitting
using supernova data, involving the most recent union 2.1 data set.
\end{abstract}

\pacs{04.50.Kd, 98.80.-k, 95.36.+x}

\date{\today}

\maketitle

%%%%%%%%%%%%%%%%%%%%%%%%%%%%%%%%%%%%%%%%%%%%%%%%%%%%%%%%%%%%%%%
%%%%%%%%%%%%%%%%%%%%%%%%%%%%%%%%%%%%%%%%%%%%%%%%%%%%%%%%%%%%%%%

\section{Introduction}

Observational evidences imply that the universe is undergoing a phase of anomalous
acceleration after a precise time, usually named the \emph{transition time}
\cite{galaxy,copeland}. In particular, the corresponding transition redshift, $z_{tr}$,
indicates at which stage the universe changed its dynamical properties and started
accelerating after a phase of deceleration \cite{peebles}. Recently, it has been argued
that constraining $z_{tr}$ provides information on the form of the fluid responsible for
the observed universe speeding up. Consequently, $z_{tr}$ may likely reveal possible new
gravitational physics due to modifications of Einstein's gravity \cite{om1}.

The fluid which triggers the current universe acceleration is often referred to as
\emph{dark energy} and fills more than the 70$\%$ of the whole universe energy budget
\cite{bamba}. The standard cosmological model assumes that the dark energy source is
supplied by the existence of a non-zero cosmological constant $\Lambda$. The corresponding
paradigm, named the $\Lambda$CDM model, is constructed by employing a net matter density
composed by baryons and cold dark matter, with a constant dark energy term
$\Omega_\Lambda\equiv {3\Lambda \over 8\pi G}$ \cite{LCDM}. Even though the model likely
represents the simplest approach for describing universe's dynamics, amongst others the
cosmological constant does not furnish an explanation to the \emph{coincidence problem}
between matter and dark energy magnitudes. In other words, since the cosmological constant
does not evolve in time, it is improbable that the ratio between matter and dark energy
densities is so close today \cite{coincidenza}. Additionally, quantum field theory
predictions forecast an enormous value for the cosmological constant if compared with the
one measured by current cosmological observations. This issue is the well known
\emph{fine-tuning} problem and represents a challenge to understand the physical origin of
the cosmological constant itself \cite{problemafine}. Due to the above caveats, one can
modify the universe content, and attribute the dark energy sector to a canonical scalar
field, a phantom field, to the combination of both fields in a unified model, or
proceed to more complicated constructions (for reviews see \cite{copeland,Cai:2009zp}).

An alternative way to reproduce the universe dynamics is by extensions of general
relativity by means of additional degrees of freedom, which do not violate the
equivalence principle, and represent a bid to formulate a semi-classical scheme for both
late and early-time universe \cite{reviewsalfel}. In the usual approach to modify
gravity, one starts by the usual curvature formulation of general relativity, and
replaces the Ricci scalar $R$ in the Einstein-Hilbert action by arbitrary functions of
it, or even more complicated curvature invariants. However, alternatively one can use as
a base the torsional formulation of general relativity, namely the so called
``teleparallel equivalent of general relativity'' \cite{ein28}, and modify its action
instead. In particular, in teleparallel gravity \cite{Maluf:2013gaa} the gravitational
field is described not by the curvature tensor but by the torsion one, and thus the
corresponding Lagrangian, namely the torsion scalar $T$, is constructed by contraction of
the torsion tensor in a similar way that in usual general relativity the Lagrangian,
namely the curvature scalar $R$, is constructed by contractions of the curvature tensor.
Hence, similarly to the $f(R)$ extension of general relativity, one can construct the
$f(T)$ extension of teleparallel equivalent of general relativity
\cite{storny1,Teleparallelism}. The interesting feature is that although general
relativity coincides completely with teleparallel equivalent of general relativity, $f(T)$
gravity is different from $f(R)$ one, thus it is a novel gravitational modification  with
rich cosmological implications \cite{Teleparallelism,storny2,storny3}.

With those considerations in mind, in this work we are interested in describing the
dark energy effects, directly calculating the corresponding transition redshift $z_{tr}$
that is predicted in $f(T)$ cosmology. In order to do so, we only consider those
$f(T)$ models which are consistent with present-time cosmographic constraints. Hence, we
aim to obtain cosmographic bounds on the $f(T)$ scenarios, by considering the modified
Friedmann equations, and then get the corresponding limits on $z_{tr}$. The main
advantage of using cosmography is that the value of $z_{tr}$ is reconstructed by means of
a model-independent procedure. Rephrasing it differently, we are able to distinguish
which classes of $f(T)$ gravity pass the cosmographic requirements and thus are viable,
by inferring the limits over $z_{tr}$ that those classes predict. In particular, we
compare this transition epoch with the one determined by the standard cosmological
paradigm, and we propose an effective cosmological model capable of reproducing the
cosmographic constraints and compatible with the limits on $z_{tr}$. Additionally, to
enable our treatment, we consider the use of the luminosity distance and we match cosmic
union 2.1 supernova data \cite{Suzuki:2011hu} with the cosmographic expansions. Thus, we
evaluate the corresponding deceleration parameter and we show that the transition redshift
is effectively comparable to the one predicted by the $\Lambda$CDM approach.

As we will see, the limits on $z_{tr}$ show that the considered $f(T)$ classes reduce to
the $\Lambda$CDM model in the lowest redshift domain, in agreement with
\cite{Nesseris:2013jea,Iorio:2012cm}. This feature indicates that the role played by the
cosmological constant may be reinterpreted as a limiting case of a more general extension,
and thus from those cosmographic corrections we show that small discrepancies occur at
$z\leq1$, whereas higher departures might be expected at high-redshift regimes. At this
point, we involve a Monte Carlo fitting procedure based on the Metropolis algorithm, in
order to compare our effective cosmological model with present-time data. Numerical
limits, priors and final outcomes, testify the efficiency of our approach, showing that
the effective torsional dark energy naturally satisfies the cosmographic requirements,
and hence it may be a candidate as a valid alternative to describe the universe dynamics.

The paper is organized as follows. In Sec. \ref{sectionII} we describe the techniques for
recovering the cosmographic settings on the $f(T)$ classes of models and we moreover
propose how to obtain $f(T)$ reconstructions. In Sec.  \ref{sectionIII} we enumerate the
properties of the transition redshift and its important role in modern cosmology.
Furthermore, we describe the main consequences in $f(T)$ gravity and we show how the
modified Friedmann equations changed when the transition occurred. In Sec. \ref{sectionIV}
we summarize the cosmographic results and we propose an effective reconstruction of $f(T)$
cosmology. To do so, we infer the deceleration parameter for the effective torsional dark
energy models and finally we show the numerical priors on the transition redshift $z_{tr}$
predicted by our paradigm. In Sec.  \ref{sectionV} we compare the cosmological
consequences of the examined models with modern data, employing the use of the union 2.1
supernova survey. We determine the free parameters of our approach and we show that the
dark energy corrections are compatible with the bounds offered by alternative dark energy
models. Finally, in Sec. \ref{sectionVI} we summarize the conclusions and perspectives of
our approach.

%%%%%%%%%%%%%%%%%%%%%%%%%%%%%%%%%%%%%%%%%%%%%%%%%%%%%%%%%%%%%%%%%%%%%%%%%%%%%%%%%%%%%%%%%%
%%%%%%%%%%%%%%%%%%%%%%%%%%%%%%%%%%%%%%%%%%%%%%%%%%%%%%%%%%%%%%%%%%%%%%%%%%%%%%%%%%%%%%%%%%

\section{The procedure for $f(T)$ reconstruction from cosmography}
\label{sectionII}

In teleparallel formulation of gravity, as well as in its $f(T)$ extension, one uses the
vierbein fields $e^\mu_A$, which form an orthonormal base for the tangent space
at each point $x^{\mu}$ defined on a generic manifold, and thus the metric reads as
$g_{\mu\nu}=\eta_{A B} e^A_\mu e^B_\nu$ (in the following greek indices
and Latin indices span the coordinate and tangent spaces respectively). Additionally,
instead of the torsionless Levi-Civita connection one uses the curvatureless
Weitzenb{\"{o}}ck one $\overset{\mathbf{w}}{\Gamma}^\lambda_{\nu\mu}\equiv e^\lambda_A\:
\partial_\mu e^A_\nu$  \cite{Maluf:2013gaa}, and therefore the gravitational field is
encoded in the torsion tensor
\begin{equation}
T^\rho_{\verb| |\mu\nu} \equiv e^\rho_A
\left( \partial_\mu e^A_\nu - \partial_\nu e^A_\mu \right).
\end{equation}
Hence, the Lagrangian of teleparallel gravity, namely the torsion scalar $T$, is
constructed by contractions of the torsion tensor as  \cite{Maluf:2013gaa}
\begin{equation}
\label{Tscalar}
T\equiv\frac{1}{4}
T^{\rho \mu \nu}
T_{\rho \mu \nu}
+\frac{1}{2}T^{\rho \mu \nu }T_{\nu \mu\rho}
-T_{\rho \mu}{}^{\rho }T^{\nu\mu}{}_{\nu}.
\end{equation}
Finally, one can extend teleparallel gravity and construct the action of $f(T)$ gravity
as \cite{storny1,Teleparallelism}
\begin{equation}
 {\mathcal S}= \int d^4x e \left[
\frac{f(T)}{2{\kappa}^2}
\right],
\label{actionfT}
\end{equation}
 where
$e= \det \left(e^A_\mu \right)=\sqrt{-g}$ and $\kappa^2$ is the gravitational
constant.

The general field equations of $f(T)$ gravity are obtained by varying the action
 ${\mathcal S}+ {\mathcal S}_m$, with ${\mathcal S}_m$ the matter action, in terms of the
vierbeins, and they read as
\begin{eqnarray}
\label{eq:IXA-2.7}
&&\!\!\!\!\!\!\!\!e^{-1}\partial_{\mu}(ee_A^{\rho}S_{\rho}{}^{\mu\nu})f^{\prime}
 +
e_A^{\rho}S_{\rho}{}^{\mu\nu}\partial_{\mu}({T})f^{\prime\prime}\ \ \ \ \  \ \ \ \  \ \
\ \ \nonumber\\
&&\ \ \ \
-f^{\prime} e_{A}^{\lambda}T^{\rho}{}_{\mu\lambda}S_{\rho}{}^{\nu\mu}+\frac{1}{4} e_ { A
} ^ {\nu}f = \frac{{\kappa}^2}{2} e_{A}^{\rho}\,{T^{(\mathrm{m})}}_\rho^{\verb| |\nu},
\end{eqnarray}
where the tensor $S_\rho^{\verb| |\mu\nu} \doteq \frac{1}{2}
\left(K^{\mu\nu}_{\verb|  |\rho}+\delta^\mu_\rho
  T^{\alpha \nu}_{\verb|  |\alpha}-\delta^\nu_\rho
 T^{\alpha \mu}_{\verb|  |\alpha}\right)$ is defined in terms of the co-torsion
$K^{\mu\nu}_{\verb| |\rho}\doteq-\frac{1}{2}\left(T^{\mu\nu}_{\verb|  |\rho} - T^{\nu
\mu}_{\verb|  |\rho} - T_\rho^{\verb| |\mu\nu}\right)$, and where ${
T^{(\mathrm{m})}}_\rho^{\verb| |\nu}$ is the energy-momentum tensor corresponding to
${\mathcal S}_m$. In (\ref{eq:IXA-2.7}) the primes denote derivatives with respect to
$T$. Finally, since for $f(T)=T$ equations (\ref{eq:IXA-2.7}) provide exactly the same
equations with general relativity, that is why the theory with $f(T)=T$ was named by
Einstein ``teleparallel equivalent of general relativity'' \cite{ein28}.

In order to apply $f(T)$ gravity in a cosmological framework we assume a spatially-flat
Friedmann-Robertson-Walker metric $ds^2=dt^2-a(t)^2\left(dr^2+r^2\,d\Omega^2\right)$,
with $d\Omega^2\doteq d\theta^2+\sin^2\theta \,d\phi^2$, which can arise from the
vierbein $e_{\mu}^A={\rm diag}(1,a,a,a)$. In this case, the field equations
(\ref{eq:IXA-2.7}) give rise to the modified Friedmann equation
\begin{subequations}
\begin{align}
\label{equazionidifriedmann1}
H^2& = \frac{1}{3} \left ( \rho_m+  \rho_{T} \right)\,,\\
    %2 \dot{\mathcal{H}} + 3\mathcal{H}^2 & = - \frac{1}{3} (p+p_{\mathcal T })\,,
    \dot{H} & = - \frac{1}{2} \left(\rho_m+p_m+\rho_T+P_{T}\right)\,,
    \label{equazionidifriedmann2}
\end{align}
\end{subequations}
with $H\doteq \frac{\dot a}{a}$ the Hubble parameter and dots indicating derivatives
with respect to the cosmic time. In the above expressions $\rho_m$ and $p_m$
are the energy density and pressure of the matter sector considered to correspond to a
perfect fluid, and moreover from now on we use units in which $\kappa^2=1$. Furthermore,
we have introduced the energy density and pressure of the effective dark
energy sector, which incorporates the torsional modifications, as
\begin{subequations}
    \begin{align}
\rho_{T}&\doteq -\frac{f}{2}-\frac{T}{2}+Tf^\prime ,
\label{rhoT}\\
\label{PT}
P_{T}&\doteq\frac{1}{2}\left[\frac{f-f^\prime T
+2T^2f^{\prime\prime}}{f+2Tf^{\prime\prime}}\right].
    \end{align}
\end{subequations}
Thus, the dark energy equation-of-state parameter writes as $w_{DE}\doteq w_T=P_T/\rho_T$.
Finally, note that for the FRW geometry, the calculation of the torsion scalar
(\ref{Tscalar}) leads to the useful relation
\begin{equation}
\label{torsioneehubble}
T=-6H^2\,.
\end{equation}

The issue of finding out a form for the dark energy equation of state passes through the
determination of the most viable forms of $f(T)$. If one knows the $f(T)$ form, it is
possible to infer the interpretations of $\rho_{T}$ and $P_{T}$ as terms associated to
torsional dark energy, i.e. a torsional contribution driving the observed cosmic
acceleration. The idea to get a viable $f(T)$ function lies on requiring that at small
redshift the $f(T)$ model reproduces the observational data and predicts a compatible
transition redshift. Hence, the strategy of this manuscript is to frame a phenomenological
reconstruction of $f(T)$ and its derivatives in terms of cosmography, which becomes a sort
of initial settings for $f(T)$ models. Having in mind these cosmographic requirements, we
simply impose the validity of the cosmological principle \cite{luongo1}, the geometrical
setting of the scalar curvature \cite{luongo2}, and the possibility of expanding $f(T)$
and its derivatives around present time in Taylor series \cite{luongo3}. We discuss below
each of those three requirements, in order to define the cosmographic series and its
application in $f(T)$ cosmology.

\begin{itemize}

\item
First, employing the cosmological principle permits to frame the universe expansion
history in terms of a single parameter, namely the scale factor $a(t)$ which enters the
Friedmann-Robertson-Walker metric as function of the cosmic time only. One gathers viable
outcomes imposing that this function may be expanded in Taylor series around present time,
and constraining the corresponding Taylor coefficients associated to the scale-factor
derivatives. This strategy compares $a(t)$'s derivatives \emph{directly} with cosmic data
and may be used as a reconstruction for the $a(t)$ shape. This benefit allows one to
distinguish among all paradigms, derived from imposing the form of $f(T)$, the ones whose
cosmographic requirements better match with data.

\item

The second caveat is the issue of spatial curvature which leads to a degeneracy problem
between its value and the variation of the acceleration. It has been proved that photon
geodesics change their paths according to its value. Thus, expanding a physical quantity
into a cosmographic series needs to fix somehow the value of spatial curvature, in order
to allow cosmography to be as model-independent as possible \cite{suggal}. According to
previous approaches \cite{luongoaltro, luongoaltrobis}, one imposes geometrical bounds on
$\Omega_k$ by assuming the matching between early and late time observations
\cite{luongoaltrotris}. We therefore assume that the universe is spatially flat, with
possible small deviations which do not influence the whole dynamics.

\item Finally, since all observable quantities of interest are assumed to smoothly evolve
as the universe expands, it is licit to assume that Taylor expansions may be easily
accounted and no saddle points or poles occur. It follows that all functions are analytic
and the cosmographic treatment is perfectly plausible \cite{vis,novis}. After those
properties, one soon expands in Taylor series the scale factor $a(t)$ as
\begin{equation}\label{adit}
a(t)\doteq \sum_{l=0}^\infty{1\over p!}a_{p}\bar{t}^{\,p}\approx
a_0\,+\,a_1\bar{t}\,+\,a_2\bar{t}\,
+\,\ldots\,,
\end{equation}
where $a_p\doteq\frac{d^pa}{dt^p}$, with $\bar{t}\doteq t-t_0$ and $t_0$ the present
time. Finally, it proves convenient to express the observable quantities under interest
in terms of the redshift $z=-1+a_0/a$.

\end{itemize}

Amongst all observables, we are much interested in the use of the luminosity distance,
since we will use supernovae Ia type to fix our cosmological bounds. Hence, imposing
$a_0=1$ in  \eqref{adit} and inserting it in the definition of the
luminosity distance:
\begin{equation}\label{defi}
D_L = (1+z)\int_{0}^{z}\frac{d\xi}{H(\xi)}\,,
\end{equation}
we write down the Taylor series around $z=0$ as
\begin{eqnarray}\label{dLTaylor}
\left\{
  \begin{array}{lll}
    D_L = \frac{z}{H_0}\,\tilde{d}_L(z;\theta)\,, & \hbox{\,} \\
    \,\\
    \tilde{d}_L(z;\theta) = 1 + d_{L1}z + d_{L2}z^2 + \ldots\,, & \hbox{\,} \\
  \end{array}
\right.
\end{eqnarray}
truncated at the third order. Moreover, applying also the definitions
\begin{subequations}
\label{Hpunto}
\begin{align}
\dot{H} =& -H^2 (1 + q)\,,\\
\ddot{H} =& H^3 (j + 3q + 2)\,,
\end{align}
\end{subequations}
we obtain the corresponding coefficients of the Taylor expansion in terms of the
cosmographic series:
\begin{subequations}\label{coefficienti}
\begin{align}
d_{L1}&=\frac{1 - q_0}{2}\,,\\
d_{L2}&=\frac{1-q_0(1+3q_{0}) +j_0 }{6}\,.
\end{align}
\end{subequations}
In these expressions we have introduced the deceleration and jerk parameters as
\begin{subequations}
\label{CSdef}
\begin{align}
q \doteq -\frac{1}{a H^2} \frac{\d^2a}{\d t^2}\,,\label{unocs}\\
j  \doteq \frac{1}{a H^3} \frac{\d^3a}{\d t^3}\,,\label{duecs}
\end{align}
\end{subequations}
indicating respectively whether the universe is accelerating or not and how the
acceleration changed sign, with the subscript ``0'' denoting the value of a quantity at
present. Observations indicate $j_0>0$ and then testify that the transition time occurred.
However, there still exists a tension between the possibilities $0<j_0<1$ and $j_0>1$
\cite{iosoltanto}.

For the sake of completeness, we  notice that it would be easy to arbitrarily extend the cosmographic series up to higher orders. For example, the next term, entering the Taylor expansion of the luminosity distance, i.e. the fourth order, would linearly depend on the snap $s\equiv \frac{1}{a H^4} \frac{\d^4a}{\d t^4}$ evaluated at present time\footnote{In general, it is straightforward to prove that any order linearly depends on the cosmographic $n$-term \cite{vissersnap}.}. However, expanding beyond the third-order leads to a non-clear physical interpretation of the corresponding coefficients. In other words, using extended cosmographic series at higher order could become quite non-predictive for our analysis, since it is difficult to physically bound coefficients beyond $j_0$. Indeed, in order to improve constraints over cosmographic coefficients at arbitrary orders, one can imagine to adopt either experimental combined tests or different definitions of cosmic distances \cite{luongo2}. In particular,  using  combined tests may generally reduce the numerical intervals of cosmographic coefficients, while any alternative distance definition would be plagued by inaccurate under/over-estimations of the cosmographic parameters, typically due to the limited cosmic available data. In addition, present time combined tests may only provide tighter ellipses where  the cosmographic coefficients span, albeit definitive constraints up to $5\sigma$ confidence level would be hardly determined.

A straightforward example of the difficulty to bound the cosmographic parameters is offered by supposing to fix the today spatial curvature  with arbitrary accuracy. In that case, it would be possible to better circumscribe the parameters beyond the third order (including jerk, snap, or higher terms)\footnote{For our purposes, we follow the strategy to take a vanishing scalar curvature to show that, under this hypothesis, the cosmographic coefficients allow compatible transition redshifts in $f(T)$ gravity.}. Unfortunately, from this procedure we can only partially exclude regions where those intervals run, getting corresponding upper and lower limits which characterize each coefficients of the Taylor expansions\footnote{Possible examples of those regions are given in \cite{refer1}. Here, the authors showed that by means of combined tests, with older supernova data sets, the jerk parameter should be positive at the $92\%$ confidence level.}. Recent analysis fixes the following limits over the cosmographic coefficients \cite{luongo1}:
\begin{subequations}\label{regioni}
\begin{align}
q_0&\in[-0.9;\,-0.4]\,,\\
j_0&\in[0.8;\,2]\,,\\
s_0&\in[-1;\,7]\,,
\end{align}
\end{subequations}
all evaluated at a $2\sigma$ confidence level. It is immediately  clear that badly constrained results can be obtained going beyond the third order of the cosmographic expansion. In particular, as stressed above, the issue associated to those regions is that they correspond to narrow (and often long) ellipses which are characterized by high error dispersions. Thus, even modern data seem to focus on strict regions for $H_0,q_0$ and $j_0$, whereas do not account for the sign of $s_0$ in the same way.

In addition, as shown in \cite{vissersnap}, one may relate cosmography to the cosmic equation of state. From this fact, it is easy to show that the second order derivative of the cosmic pressure with respect to the total density linearly depends upon $s_0$ itself. Alternatively, even the second derivative of the acceleration parameter, i.e. $\frac{d^2q}{dz^2}$, cannot be accurately constrained with current numerics on $s_0$ reported in \eqref{regioni}.

Summarizing, adopting present data, all cosmographic quantities suffer from badly bounded numerical results, so that the  use of $s_0$ in our approach would influence the experimental analysis itself, providing broadening systematics and higher dispersions in the evaluations of the transition redshift $z_{t}$. For those reasons, in order to consider fourth order expansions, one should adopt improved intervals of cosmic data or using   distance definitions that do not somehow depend on scalar curvature, i.e. where the scalar curvature is not fixed as a prior  coming from  other observations. Hence, in this paper, we definitively use a truncated third order Taylor expansions of $D_L$ thanks to the above considerations. In so doing, the numerical results of $H_0,q_0$ and $j_0$  lead to acceptable dispersions and allow us to better circumscribe the intervals for $z_{tr}$ in a more suitable way.

Thus, from the above analysis it follows that the cosmographic series is the set of coefficients
evaluated at present time. The cosmographic series has been built up in function of a
Taylor series expanded around $z=0$, albeit it is possible to handle it also in terms of
the cosmic time. This is clearly possible involving the definition of the redshift in
terms of the cosmic time as:
\begin{equation}
\label{cosmoz}
\frac{\d { z} }{(1+z)}=-H(z)\d t\,.
\end{equation}

Since the cosmographic series may be expressed in terms of the scale-factor derivatives,
it does not depend upon the particular choice of the cosmological model. This property
represents a key to conclude that imposing a cosmological model \emph{a priori} is
unnecessary and any modified gravity may be limited by assuming the cosmographic
requirements. We here follow the technique of reconstructing the $f(T)$ models by means of
late-time cosmography. To do so, we rewrite the luminosity distance \eqref{defi} in terms
of $f(T)$ derivatives. This is possible since there exists a direct correspondence between
$q_0$ and $j_0$ with the $f(T)$ form and its derivatives. In particular, rewriting
\eqref{coefficienti} as function of $f(T)$, we frame the effective model derived from the
torsional dark energy by directly comparing $D_L$ with data. Rephrasing it differently,
instead of using $q,j,\ldots$ we consider $f(T), f^\prime(T), f^{''}(T), \ldots$ and we
obtain the numerical outcomes on those quantities. The cosmographic constraints on $f(T)$
and its derivatives, point out the numerical priors that we use as \emph{initial
settings} for reconstructing the shape of viable effective $f(T)$ models, which reproduce
dark energy at small redshift. Afterwards, we predict the transition redshift from our
effective model and we understand whether our model indicates a viable $z_{tr}$ if
compared with the $\Lambda$CDM predictions. We will report the connections between the
cosmographic series and the $f(T)$ derivatives in Sec. IV.

%%%%%%%%%%%%%%%%%%%%%%%%%%%%%%%%%%%%%%%%%%%%%%%%%%%%%%%%%%%%%%%%%%%%%%%%%%%%%%%%%%%%%%%%%%
%%%%%%%%%%%%%%%%%%%%%%%%%%%%%%%%%%%%%%%%%%%%%%%%%%%%%%%%%%%%%%%%%%%%%%%%%%%%%%%%%%%%%%%%%%

\section{The $f(T)$ transition redshift}
\label{sectionIII}

In the scenario at hand, the $f(T)$ term drives the dark energy contribution, interpreting
the dark sector as due to a \emph{torsional dark energy}. It is widely believed that the
dark energy contribution dominates over matter at our time, while it appears negligible at
higher redshift regimes. The type of the transition and the time at which it occurs are
extremely relevant, since they indicate the dark energy nature and may also provide
information on how the dark energy evolves in time. In particular, the transition time and
correspondingly the transition redshift emphasize the change from decelerated to
accelerated cosmological expansion, and represent a prediction of any particular model
involved to describe the universe expansion history. In other words, direct measurements
of the transition redshift provide direct information on both the deceleration and
acceleration epochs.

To show how to investigate $z_{tr}$ in the framework of $f(T)$ gravity, let us consider
the definition of the transition redshift, which occurs at a zero of the deceleration
parameter $q$. We will find out the transition redshift $z_{tr}$ for a class of
cosmographic $f(T)$ models, and we will also compare it with standard model
predictions and with recent bounds on $z_{tr}$ itself.

Passing through the phase of transition between matter and dark energy dominance, and
assuming the matter to be dust (i.e. $P_m=0$), it is useful to combine the two Friedmann
equations (\ref{equazionidifriedmann1}), (\ref{equazionidifriedmann2}) to obtain the
torsional pressure in terms of the deceleration parameter as:
\begin{equation}
\label{eqassunta}
P_T=H^2(2q-1)\,,
\end{equation}
where we made use of relation \eqref{unocs}. At the transition time we therefore obtain
the value of the torsional pressure as
\begin{equation}
\label{rtyu}
P_T(z_{tr})=-H^2_{tr}\,,
\end{equation}
which corresponds to $q=0$ at the transition redshift $z_{tr}$, with Hubble rate
$H_{tr}$. This expression is equivalent to the standard barotropic dark-energy pressure in
the framework of general relativity given by \cite{sterano}:
\begin{equation}
\label{poil}
P_{tr}=-H^2_{tr}\,.
\end{equation}
In particular, the two results, \eqref{rtyu} and \eqref{poil}, lead to the same formal
outcome. In fact, assuming $H_{tr}$ to be positive definite, both the torsional and
standard dark-energy pressures are negative at the transition. However, the physical
meaning behind \eqref{rtyu} and \eqref{poil} is different, in the sense that in the first
case the transition is induced by the torsional terms, while in the standard approach the
transition is realized due to the dark energy or curvature terms.

In the standard  $\Lambda$CDM cosmological model, the cosmological constant contributes
about 70$\%$ of the present cosmological energy budget and the consequence on cosmology
lies on an evolving deceleration parameter $q$ of the form \cite{galaxy}
\begin{equation}
\label{dkjnh}
q_{\Lambda}=-1+\frac{3\Omega_{m,0}(1+z)^3}{2 + 2 \Omega_{m,0} z[3+z(3+z)]}\,,
\end{equation}
with $\Omega_{m,0}$ the present matter density parameter, and where the corresponding
transition redshift formally is given by
\begin{equation}
\label{dn}
z_{tr}=\Big(\frac{1}{H}\frac{dH}{dz}\Big)^{-1}\Big|_{z=z_{\mathrm{tr}}}-1\,,
\end{equation}
which has been obtained assuming $\ddot a=0$. Thus, the $\Lambda$CDM model gives an exact
solution for $z_{tr}$, of the form
\begin{equation}
\label{trans}
z_{tr,\Lambda}=\left[2\frac{(1-\Omega_{m,0})}{\Omega_{m,0}}\right]^{1/3}-1\,.
\end{equation}

In the following section we will use relations \eqref{eqassunta} and \eqref{rtyu} in
order to infer the numerical values of the torsional pressure at the transition time.
Afterwards, we will quantify the difference of the standard predictions of the
$\Lambda$CDM model with those obtained in the present cosmographic approach. We will
therefore predict $z_{tr}$ for $f(T)$ cosmology and we will compare it with expression
\eqref{trans}.

%%%%%%%%%%%%%%%%%

\section{Reconstructing effective cosmographic $f(T)$ models}
\label{sectionIV}

Let us now apply the approach described in the previous two sections and proceed to the
reconstruction of effective cosmographic $f(T)$ models. We start by using
(\ref{torsioneehubble}), as well as  \eqref{unocs},\eqref{duecs}, in order to calculate
the time derivatives of $T$ as
\begin{subequations}\label{cosmoto}
\begin{align}
  \dot{ T} &= 12\, H^3\,(1+q)\,, \\
  \ddot{T} &= -12\,H^4\,(q^2+j+5q+3)\,,
\end{align}
\end{subequations}
with the connection between the torsion and $P_T$ reading as
\begin{equation}\label{condizionesuT}
T=\frac{6P_T}{1-2q}\,.
\end{equation}
Hence, at the transition time, we have
\begin{eqnarray}
%  T&=&6H_{tr}^2=2P_{T,tr}\\
  \dot{T} &=& 12\, H^{3}_{tr}\,,\\%=12\,3^{-{3\over2}}P_T^{{3\over2}}\,, \\
  \ddot{T} &=& -12\,H^{4}_{tr}\,(j_{tr}+3)\,.
\end{eqnarray}
The cosmographic requirements give the connection between $f,\,f^\prime$ and $f''$ from
the modified Friedmann equations, namely \cite{luongo3}
\begin{subequations}
\label{normali}
\begin{align}
f( T_0)&=6 H_0^2\,(\Omega_{m,0}-2)\,,\\
f''( T_0)&=\frac{1}{6 H_0^2}\Big[\frac{1}{2}-\frac{3\Omega_{m,0}}{4(1+q_0)}\Big]\,,
\end{align}
\end{subequations}
and moreover $f'(T_0) = 1$ to guarantee the Solar-System constraints in order to
preserve the value of $G$ at our time \cite{luongo3}.

Thus, using expressions \eqref{cosmoto} and \eqref{normali}, we acquire the priors on
$T$, $f(T)$ and $f''(T)$ as
\begin{subequations}
\label{no}
\begin{align}
f(T_0)&\in[-5.23;\,-4.79]\,,\\
f''(T_0)&\in[-0.19;\,0.01]\,,\\
T_0&\in[-3.11;\,-2.77]\,,\\
\dot{T} &\in[0.75;\,2.24]\,, \\
\ddot{T} &\in[-7.26;\,-0.36]\,,
\end{align}
\end{subequations}
which have been obtained assuming a normalized Hubble rate $H_0\in[0.68;\,0.72]$ and a
mass density $\Omega_{m,0}\in[0.274;\,0.318]$, with $q_{0}\in[-0.8,-0.5]$,
$j_0\in[0.5,1.5]$ and $j_{tr}\approx j_0$ \cite{planck}. The last condition has been
imposed assuming that the universe is slightly evolving in the redshift domain $z\leq1$.
We stress that the priors (\ref{no}) are the requirements that determine whether
a specific $f(T)$ form is viable or not.

The strategy is the following: we assume the validity of the cosmographic series as the
initial conditions of the modified Friedmann equations, and then we integrate the
first Friedmann equation. Thus, we infer the numerical values of $H(z)$ for different
redshifts, and we separately extrapolate those points, determining a list of numbers
for $H(z)$ and $z$. Finally, through the use of testing functions, we reconstruct an
effective $f(T)$ which reproduces the numerical limits. Hence, from this function one
obtains a parameterized cosmological model, which departs from the $\Lambda$CDM scenario,
corresponding to a varying dark-energy sector.

Our treatment suggests that a possible approximation of the dark-energy density term
$\rho_{DE}$ may be
\begin{equation}
\label{darkenergysolo}
\rho_{DE}\approx \log\Big[\alpha+\beta\,\sum_{i=0}^{\mathcal{N}}a^{i}\Big]\,.
\end{equation}
Truncating at the second order in $a$, we obtain the Hubble rate as
\begin{equation}
\label{Heffettivo}
\frac{H^2}{H_0^2}= \Omega_m(z) + \log[\alpha+\beta(2-3a+a^2)]\,,
\end{equation}
where we considered $\Omega_{m}\equiv\Omega_{m,0}(1+z)^3$. The parameter
$\alpha$ is fixed in order to guarantee that at $z=0$ the Hubble rate is identically
$H=H_0$. Therefore, we have
\begin{equation}
\label{valorealfa}
\alpha=e^{1-\Omega_{m,0}}\,.
\end{equation}
Hence, the cosmographic reconstruction of torsional dark energy provides a deceleration
parameter of the form
{\small{
\begin{equation}
\label{qdicou}
q=\frac{3\Omega_{m,0}(1+z)^5\alpha+\beta+3z[1+\Omega_m(z)(1+2z)]\beta}{2\Big[(1+z)^2
\alpha+z(1+2z)\beta\Big]\Big[\Omega_m(z)+\log\Big(\frac{\alpha+\beta
z(1+2z)}{(1+z)^2}\Big)\Big]}-1.
\end{equation}}}

\begin{figure}[ht]
\includegraphics[scale=0.6]{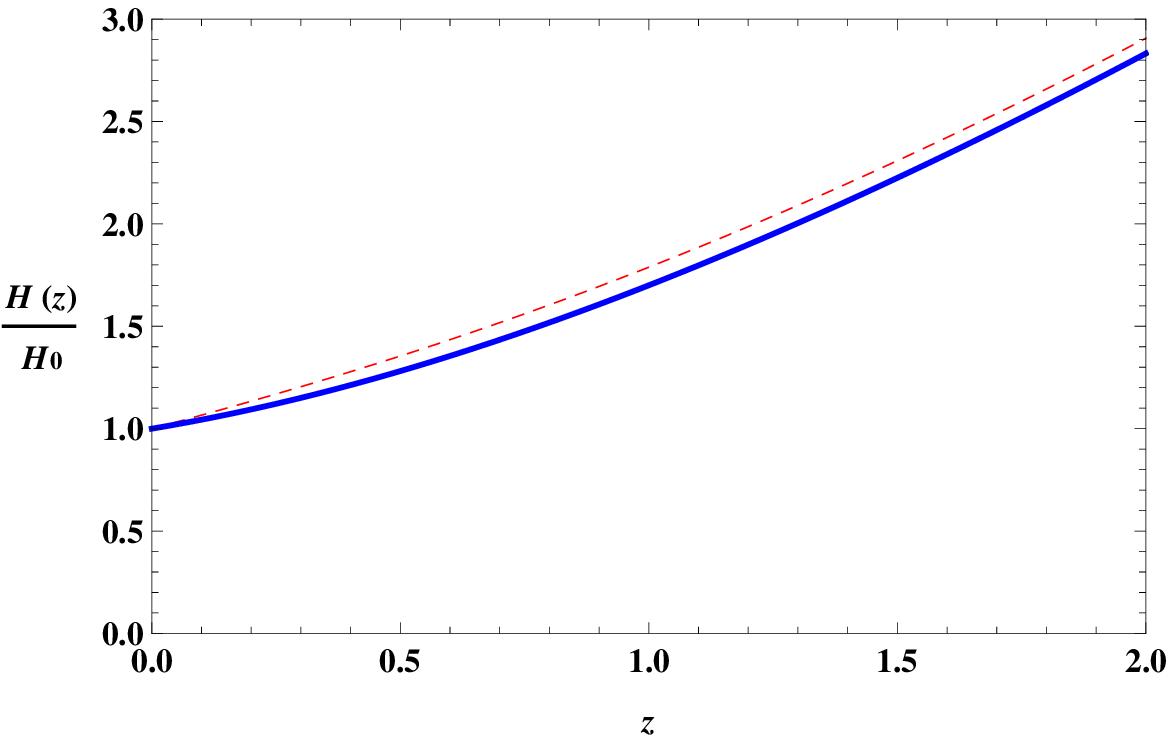}
\includegraphics[scale=0.6]{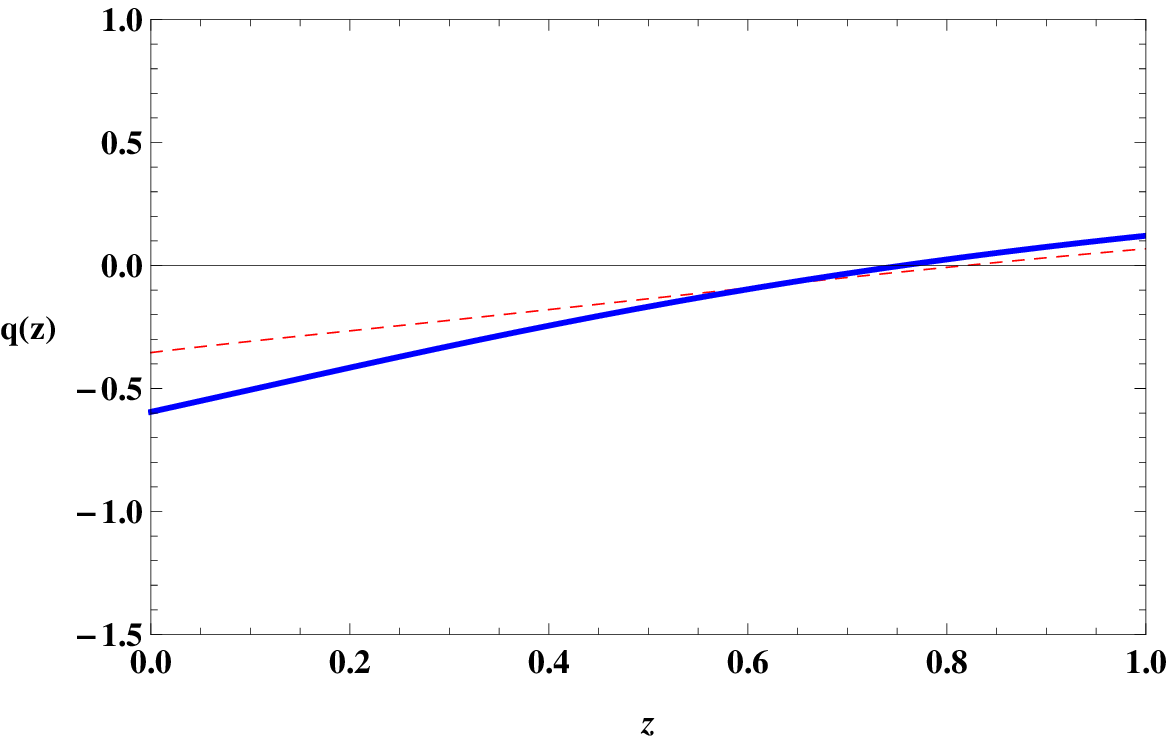}
\caption{{\it{The evolution of $H(z)/H_0$ (upper graph) and $q(z)$ (lower graph),
according to viable $f(T)$ cosmology (blue-solid curves) versus the $\Lambda$CDM
predictions (red-dashed curves). We employed the indicative values
$\Omega_{m,0}=0.27$ and $\beta=1$ and we normalized through $H_0=100Km\,Mpc^{-1}\,s$.}}}
\label{figura2}
\end{figure}

\begin{figure}
\includegraphics[scale=0.6]{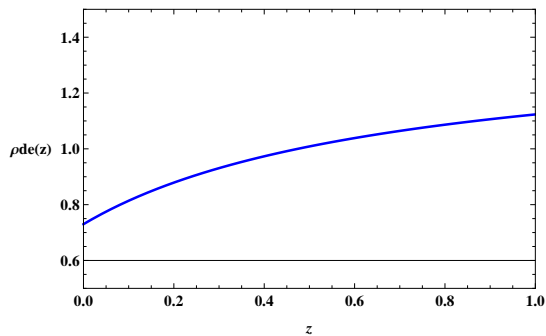}
\caption{
{\it{The evolution of the dark-energy density according to viable $f(T)$ cosmology
(blue-solid curves) versus the $\Lambda$CDM
predictions (red-dashed curves). The two cosmological paradigms exhibit very similar
behaviors, and thus the $\Lambda$CDM curve is almost indistinguishable from the viable
$f(T)$ one. We employed the indicative values
$\Omega_{m,0}=0.27$ and $\beta=1$ and we normalized through $H_0=100Km\,Mpc^{-1}\,s$.}}}
\label{figura3}
\end{figure}

In Fig. \ref{figura2} we depict the behaviors of $H(z)/H_0$ and $q(z)$, given in
\eqref{Heffettivo} and \eqref{qdicou} respectively. We deduce that up to the redshift
domain $z\leq2$, our approach is compatible with the standard cosmological model, and in
fact only small differences occur between our predictions and the
$\Lambda$CDM ones, which are slightly larger. This is due to the fact that our $H(z)$
parameter indicates a dark-energy evolution which does not departure significantly from
the case of a constant dark-energy term at small redshifts. Hence, our Hubble rate well
approximates the standard $\Lambda$CDM contribution, slightly evolving as the redshift
increases. This is more evident in Fig. \ref{figura3}, in which we plot the dark-energy
term \eqref{darkenergysolo}, normalized by means of the standard critical density
$\rho_c\equiv\frac{3H_0^2}{8\pi G}$.

Afterwards, linearizing the deceleration parameter around $z=0$, keeping first-order
terms, we
find the transition redshift as
%\begin{widetext}
\begin{equation}
\label{zetatrans}
\!z_{tr}
\simeq \frac{\alpha^2 (\Omega_{m,0} + \log\alpha)
(2\log\alpha-\Omega_{m,0}-\beta)}{[9\Omega_{m,0}
\alpha^2+(\alpha-\beta)\beta]\log\alpha-\beta [\beta + \Omega_{m,0}(5\alpha+\beta)]}.
\end{equation}
%\end{widetext}
 We mention that this approximation is efficient, since one expects a
transition at $z\leq1$ and therefore the linearized $q$ does not
substantially differ from the exact value.

Having in mind the form of the Hubble rate, we can infer limits over $\Omega_{m,0}$ and
$\beta$. This permits one to determine $z_{tr}$ from expression \eqref{zetatrans}. In the
next section we describe the fitting procedure using supernova data, and we extract
numerical bounds on the free parameters of our cosmographic torsional dark-energy
scenario.

\section{The matching with observations}
\label{sectionV}

The above approach provided a particular set of cosmographic quantities related to the
$f(T)$ form. Correspondingly, the effective Hubble rate was built in terms of corrections
to the simple teleparallel gravity, that is to general relativity. Hence, these
corrections are due to the difference between $f(T)$ cosmology and the standard paradigm
of $\Lambda$CDM cosmology. These terms may be compared with the cosmological constant
value without showing great departures at small redshift regimes, as we discussed in the
above section. The contribution due to the $f(T)$ sector needs to match an adequate
convergence at small and high redshift domains. Hence, it is required to obtain bounds on
the observational parameters, in order to understand whether the cosmological model at
hand passes or not the observational constraints.

In order to proceed, we perform the analysis by involving the Monte Carlo
technique with the use of the union 2.1 supernova compilation ~\cite{Suzuki:2011hu}. This
survey is built up by 580 measurements of apparent magnitudes, with the corresponding
redshifts and magnitude errors. Assuming a Gaussian distribution, one acquires the
relevant fact that the luminosity distance may be rewritten in terms of the cosmographic
series itself. Thus, all observations may be performed by directly fitting $D_L$ with
union 2.1 data.

We employ type Ia supernova observations since they probably represent the most suitable
cosmic compilation. The role of supernovae has been crucial for cosmological
parameter-fittings, since supernovae are considered standard candles. It follows
that their luminosity curves are easily related to distances themselves\footnote{Frequently, cosmologists showed that when different light curve fitters are used, different results with significative discrepancies may be found. For the case of $f(T)$ gravity see for example \cite{bengo}}.

The union 2.1 data set is capable of reducing previous systematics, entered in old
catalogs, for instance in union and union 2 \cite{Amanullah:2010vv,Kowalski:2008ez}.
Hence, it is easy to show that one can use the well known {\it $\chi$-squared} function,
which is commonly involved to quantify theoretical and observational distance modulus. In
particular, one defines it as \cite{GoliathAmanullah}
\begin{equation}
\chi^2 = \mathcal A - \frac{\mathcal B^2}{\mathcal C} + \mathcal D\,,
\end{equation}
where
\begin{eqnarray}
 \mathcal A &=& {\bf x}^T C^{-1} {\bf x}\,, \nonumber \\
 \mathcal B &=& \sum_i (C^{-1}{\bf x})_i\,,  \\
 \mathcal C &=& \text{Tr}[\,C^{-1}\,]\,,  \nonumber\\
 \mathcal D &=& \log \left( \frac{\mathcal C}{2\pi}\right)\,,\nonumber
\end{eqnarray}
with $C$ the covariance matrix of observational data, and where  ${\bf x}$ is
\begin{equation}
 {\bf x}_i= 5\log_{10}\left[\frac{D_L(z_i; { \theta})}{\text{Mpc}}\right] + 25 -
\mu_{obs}(z_i)\,.
\end{equation}
In particular, having a spatially flat universe, the luminosity distance simply reduces to
\begin{equation}\label{dl}
d_L={1\over a}\int_{0}^{\psi}\frac{d\psi}{H(\psi)}\,.
\end{equation}
The Hubble derivatives, evaluated as a function of the redshift $z$  at our time ($z=0$),
give us
\begin{subequations}
\label{derivate}
\begin{align}
\frac{dH}{dz}&=H_0\,\left(1+q_0\right)\,,\\
\frac{d^2H}{dz^2}&=H_0\,\left(j_0 - q_0^2\right)\,,\\
\end{align}
\end{subequations}
providing a third order Taylor series for the luminosity distance of the form: $\tilde
d_L= \eta_1\,
z+\eta_2 z^2+\eta_3 z^3 + \ldots$, where
\begin{subequations}
\label{alfabetagamma}
\begin{align}
\eta_1&=1\,,\\
\eta_2&=2-\frac{3\Omega_{m,0}+\beta\exp\left(\Omega_{m,0}-1\right)}{2}\,,\\
\eta_3&=\frac{1}{8}
\Big[3\Omega_{m,0}+\beta\exp\left(\Omega_{m,0}-1\right)\Big]\nonumber\\
&\ \ \  \ \cdot \Big[ 3\Omega_ { m ,
0}-2+\beta\exp\left(\Omega_{m,0}-1\right)\Big] \,.
\end{align}
\end{subequations}
It is useful to stress here that expression (\ref{dLTaylor}) represents a general Taylor
expansion and may be applied to any cosmological model. The advantage of passing through
it is that one directly fits a particular model of interest, weighting the coefficients
\emph{directly} with the most recent data. A simple strategy for definitively
alleviating the problem of matching data with our model, is to assume \emph{a priori}
compatible cosmographic priors. To do so, we employ the theoretical bounds given by
\eqref{no}.

This treatment represents the key to obtain suitable cosmographic intervals, in which the
free parameters of our model, i.e. $\Omega_{m,0}$ and $\beta$, do not violate the
cosmological limits. Our numerical outcomes also need to be compatible with the ones
already proposed in the literature and do not have to influence the analyses themselves.
In addition, we aim at finding out numerical outcomes over $z_{tr}$, which can be
indirectly derived from the experimental analysis by using \eqref{zetatrans}.

Hence, the Bayesian technique provides the likelihood function:
\begin{equation}
\label{jhfdkjdf}
    \mathcal{L}
\propto \exp (-\chi^2/2 )\,,
\end{equation}
whose maximum corresponds to the minimum of the $\chi^2$. We obtain our numerical results
performing a test with the free available code ROOT and the additional package BAT
\cite{programmi}. Our analyses are based on two statistical treatments, characterized by
different maximum order of parameters. We perform such a procedure in order to provide a
hierarchy among all parameters. Firstly, we allow all parameters to freely vary
(Fit1), and secondly we fix the mass density parameter through values compatible with
the most recent Planck measurements \cite{planck} (Fit2).

Our numerical results are summarized in Table \ref{tabella}, where we separately report
the obtained and the inferred results, showing the limits on the transition redshift
itself.

%%%%%%%%%%%%%%%%%%%%%%%%%%%%%%%%%%%%%%%%%%%%%%%%%%%%%%%%%%%%%%%%%%%%%%%%%%%%%%%%%%%%%%%%%%

\begin{table}[ht]
\begin{tabular}{c|c|c}

\hline\hline

{\small $\quad$ Parameter $\quad$}  &   {\small $\qquad$ Fit1 $\qquad$ }    &
{\small $\qquad$ Fit2 $\qquad$ }\\

\hline

{\small$H_0$}       & {\small $ 69.490$}{\tiny ${}_{ - 0.379}^{ +  0.366}$}        &
{\small $69.
450$}{\tiny ${}_{ - 0.355}^{ +  0.342}$}\\[0.8ex]

{\small$\alpha$}       & {\small $1.367$}{\tiny ${}_{-0.252}^{+0.296}$}      & {\small
$2.067$}{\tiny ${}_{---}^{---}$}\\[0.8ex]

{\small$\beta$}       & {\small $-1.147$}{\tiny ${}_{ - 0.502}^{ +  0.696}$}         &
{\small $0.
834$}{\tiny ${}_{ - 0.168}^{ +  0.186}$}\\[0.8ex]

{\small$\Omega_{m,0}$}       & {\small $0.687$}{\tiny ${}_{ - 0.185}^{ +  0.217}$}
& {\small $0.274$}{\tiny ${}_{---}^{---}$}\\[0.8ex]

{\small$z_{tr}$}       & {\small $0.247$}{\tiny ${}_{ - 0.271}^{ + 0.345}$}         &
{\small $0.
643$}{\tiny${}_{-0.030}^{+0.034}$}\\[0.8ex]

{\small$\Big|\Delta z_{tr}\Big|$}       & {\small $0.385$}{\tiny ${}_{---}^{---}$}
& {\small $0.011$}{\tiny ${}_{---}^{---}$}\\[0.8ex]

{\small$P_{T,tr}$}       & {\small $-0.687$}{\tiny ${}_{-0.641}^{+0.660}$}       & {\small
$-1.032$}{\tiny ${}_{-0.063}^{+0.070}$}\\[0.8ex]

%{\small$\Delta P_{T,tr}$}       & {\small $1.591$}{\tiny ${}_{-5.244}^{+6.493}$}
%& {\small $8.559$}{\tiny ${}_{-0.441}^{+0.492}$}\\[0.8ex]

%{\small$T_{T,tr}$}       & {\small $...$}{\tiny ${}_{-3.198}^{+3.909}$}         & {\small
%$...$}{\
%tiny ${}_{-6.469}^{+10.905}$}\\[0.8ex]

%{\small$\dot T_{T,tr}$}       & {\small $...$}{\tiny ${}_{-3.198}^{+3.909}$}         &
%{\small $...
%$}{\tiny ${}_{-6.469}^{+10.905}$}\\[0.8ex]

%{\small$\ddot T_{T,tr}$}       & {\small $...$}{\tiny ${}_{-3.198}^{+3.909}$}         &
%{\small $..
%.$}{\tiny ${}_{-6.469}^{+10.905}$}\\[0.8ex]
\hline \hline
\end{tabular}
\caption{\label{tabella} Table of our experimental and \emph{a posteriori}-derived
results. We report the 1$\sigma$ confidence level errors for our fitting procedure,
performed through the Metropolis algorithm. The associated errors on derived quantities
have been obtained through the logarithmic rule \cite{galaxy,pytha}. To evaluate the transition
redshift in the $\Lambda$CDM model we considered $\Omega_{m,0}=0.315$ from the Planck
measurements. Finally, $H_0$ is given in Km/s/Mpc.}
\end{table}
%%%%%%%%%%%%%%%%%%%%%%%%%%%%%%%%%%%%%%%%%%%%%%%%%%%%%%%%%%%%%%%%%%%%%%%%%%%%%%%%%%%%%%%%%%

\begin{figure}
\includegraphics[scale=0.45]{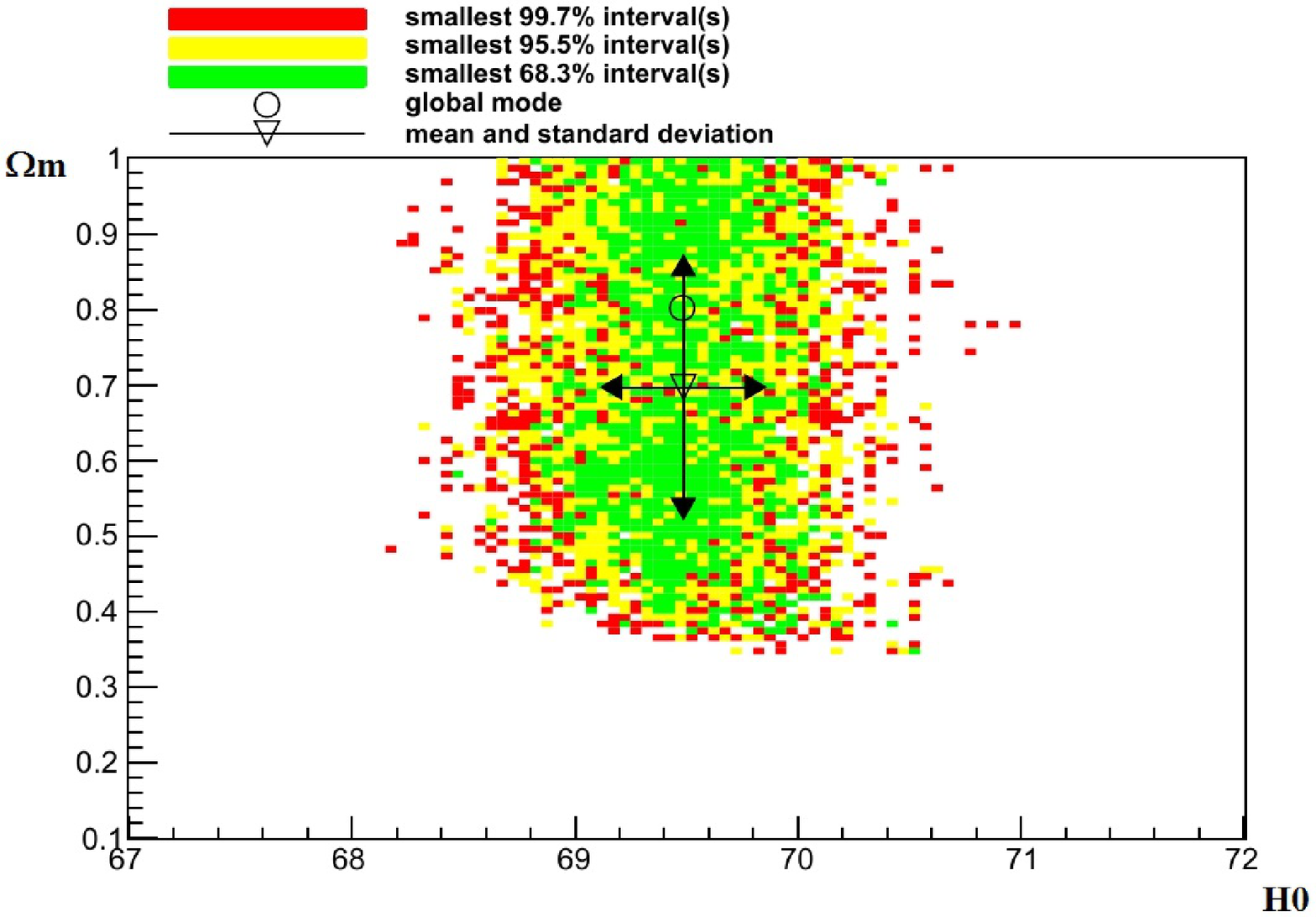}
\includegraphics[scale=0.45]{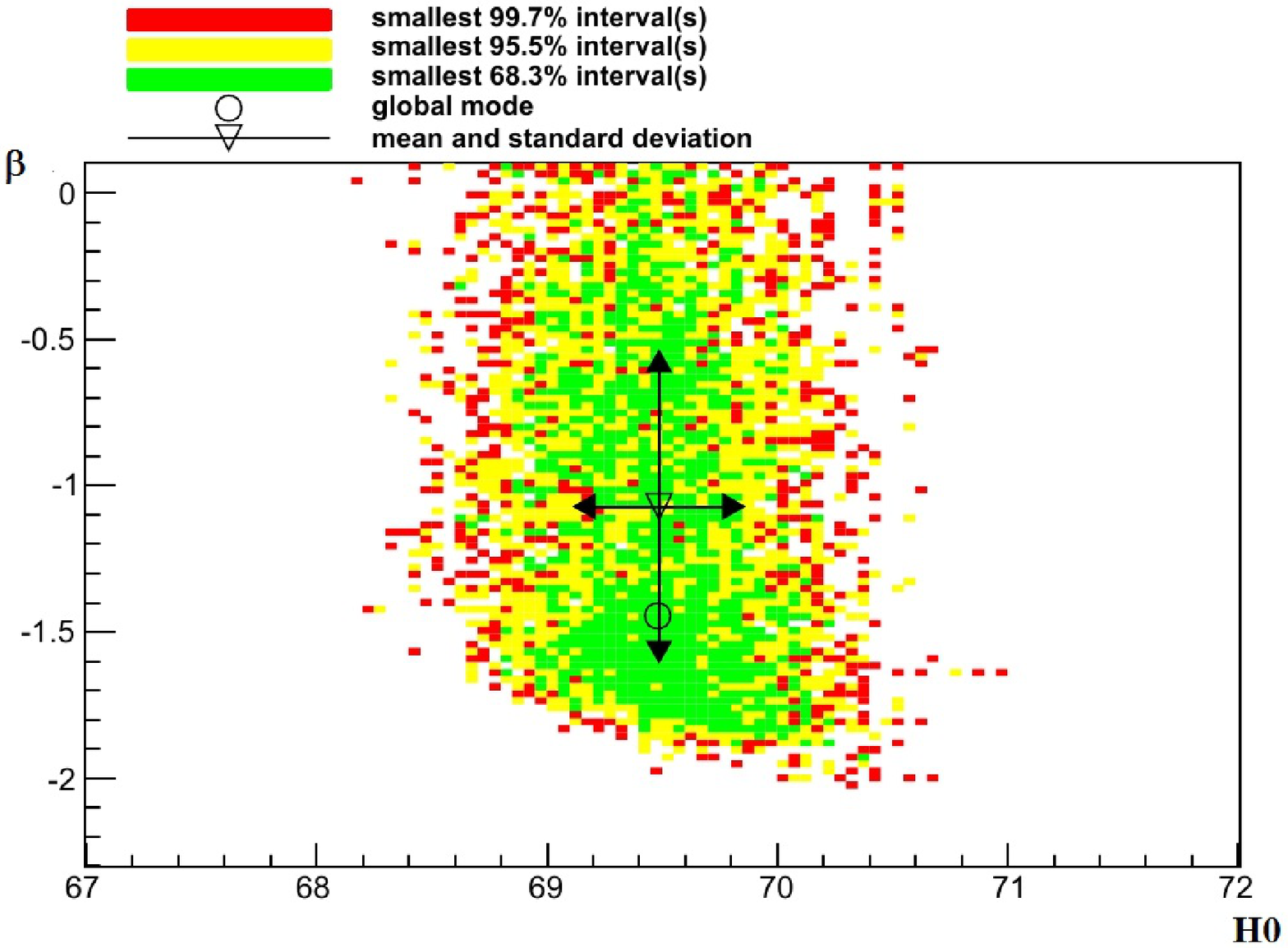}
\includegraphics[scale=0.45]{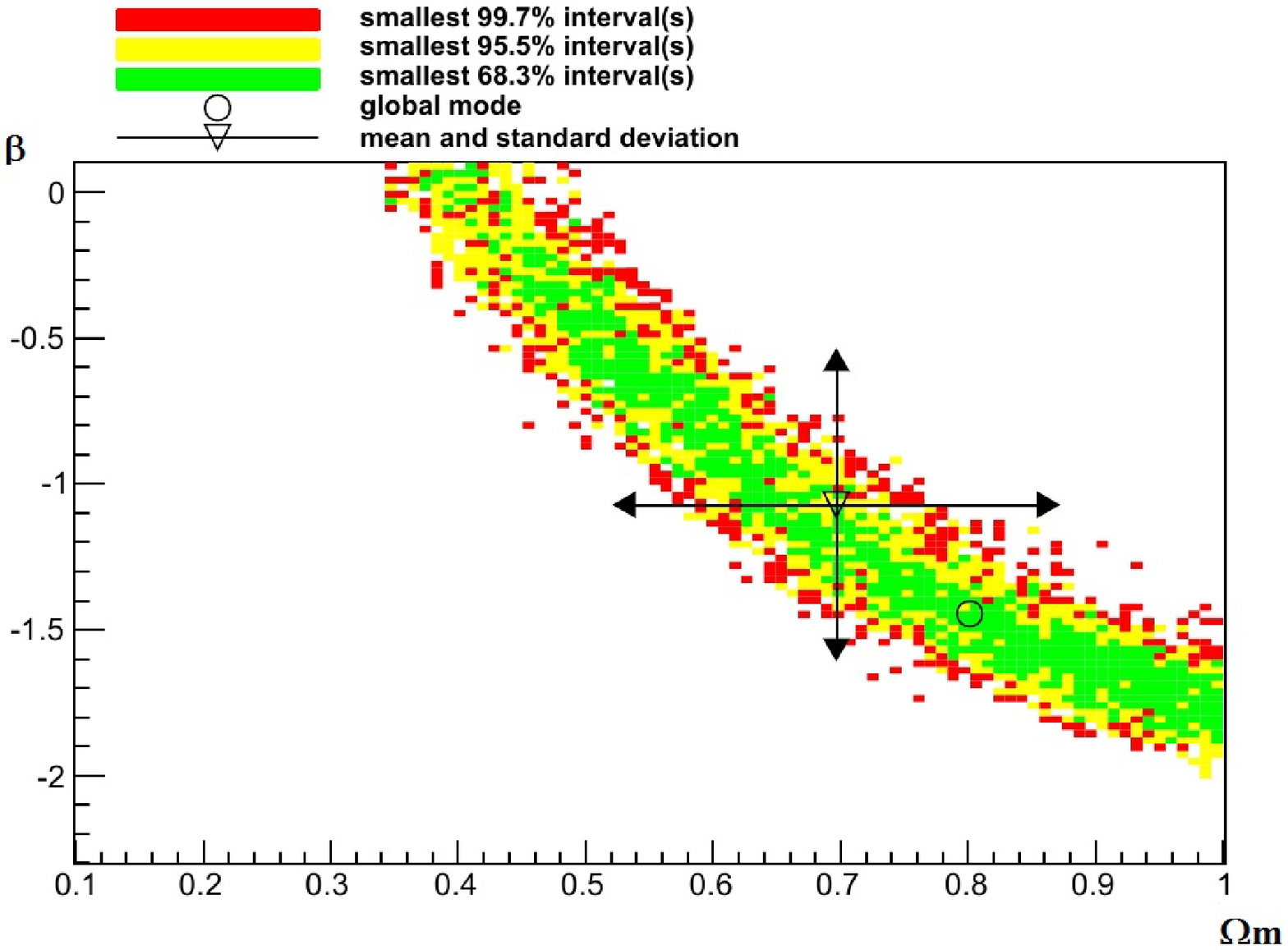}
\caption{{\it{Contour plots for our observational analyses in the case where all
parameters are free to vary (Fit1):  $\Omega_{m,0}$ versus  $H_0$ (upper graph), $\beta$
parameter versus $H_0$ (middle graph) and  $\beta$  versus $\Omega_{m,0}$ (lower graph).
}}}
\label{figura1a}
\end{figure}

The cosmological results show that the first fit (Fit1), in which all coefficients are
taken free, does not give conclusive results. In this case, in fact, the mass density is
overestimated probably due to the strong multiplicative degeneracy between the
coefficients $\Omega_{m,0}$ and $\beta$, as one can see from  \eqref{alfabetagamma}.
The cosmographic analysis suffers from this kind of degeneracy and shows the same
inefficiency in bounding $z_{tr}$, which seems to significantly departure from the
$\Lambda$CDM predictions, as shown by looking at the $\Delta z_{tr}$. The
likelihood contours of this case are shown in Fig. \ref{figura1a}.

\begin{figure}
\includegraphics[scale=0.46]{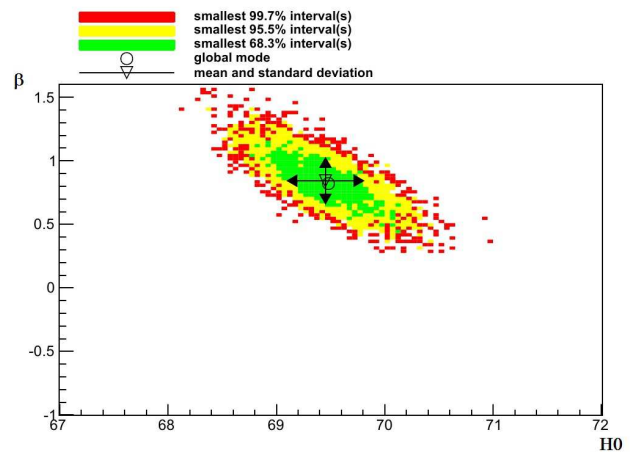}
\caption{{\it{ Contour plots for our observational analyses in the case where only $H_0$
and $\beta$ are free to vary, and $\Omega_{m,0}$ is fixed to a value compatible with the
Planck measurements, namely $\Omega_{m,0}=0.274$ (Fit2):  $\beta$ parameter versus
$H_0$.}}}
\label{figura1b}
\end{figure}

In the second fit (Fit2), where we fix the matter density parameter to a value compatible
with the Planck measurements, namely $\Omega_{m,0}=0.274$ \cite{planck}, the results are
mostly accurate. This fixing enables to get refined limits even on the other two free
coefficients of our model, as can be seen in Fig. \ref{figura1b} (compare with the middle
graph of Fig. \ref{figura1a}). As a consequence, we obtain more precise bounds on
$z_{tr}$, which becomes perfectly compatible with the constraints predicted by the
$\Lambda$CDM model, at the $1-\sigma$ confidence level. Our value however seems to be
slightly smaller than theoretical expectations ($z_{tr}=0.74$ according to \cite{lcdmtr}).

Hence, we conclude that combined observational tests will represent a landscape to better
fix constraints over the involved quantities, showing more accurate limits on the
transition predicted by $f(T)$ gravity. Further, this seems to be evident by looking at
the contour plots of Figs. \ref{figura1a} and \ref{figura1b}, in which a delineated curve
corresponds to the second fit. In other words, from these Figures it is clear that
the first fit shows higher errors since the contours are larger than the one inferred from
the second fit. Nevertheless, in all cases the predicted transition time occurs at $z<1$,
in agreement with the standard theoretical framework.

\section{Conclusions and perspectives}
\label{sectionVI}

In this paper we investigated the transition redshift derived in an effective model
inferred from $f(T)$ gravity. In order to do so, we extracted an approximate
reconstruction of the $f(T)$ dark-energy term. The effective dark energy contribution has
been obtained by numerically solving the Friedmann equations, employing as initial
conditions the numerical outcomes obtained from cosmographic bounds. In this way, we
defined a set of numerical constraints on $f(T)$ and its derivatives in a
model-independent way, and we were able to fix the evolving dark-energy term through a
logarithmic correction.

Our cosmographic model well adapts to the late-time constraints, and it reproduces a
cosmological model which smoothly departures from the standard $\Lambda$CDM paradigm. The
corresponding limits on the free parameters of the model have been obtained by directly
fitting the luminosity distance with supernova data, using the most recent union 2.1
compilation. We extracted viable constraints on the free parameters of the scenario, in
two distinct fits with different hierarchy between coefficients. We first considered all
parameters free to vary and afterwards we fixed the value of the matter density
consistently with current Planck results. All predictions provided intervals for the
transition redshift which are compatible with present expectations, although the numerical
outcomes are slightly smaller than the ones predicted by the standard cosmological model.
Departures have been encountered in the case where we leave all parameters free to vary,
due to the degeneracy problem between coefficients in the luminosity distance definition.
Possible approaches will be devoted to better fix those constraints by means of combined
cosmological tests. Moreover, we could mostly investigate the properties of our
logarithmic corrections, studying their consequences in the early phases of the
universe evolution.

Finally, it would be interesting to extend the above analysis in the case of higher-order
torsional cosmology, and in particular in the case where the teleparallel equivalent of
the Gauss-Bonnet combination is used in the action, as in $f(T,T_{\cal G})$ cosmology
\cite{Kofinas:2014owa}. The corresponding results could be compared with both
$\Lambda$CDM cosmology, as well as with the $f(R,{\cal G})$ cosmology \cite{DeFelice:2010sh,felix,antonio}.
Such an analysis could provide more information on the possible distinguishability of
curvature and torsional gravity using cosmographic methods.

%%%%%%%%%%%%%%%%%%%%%%%%%%%%%%%%%%%%%%%%%%%%%%%%%%%%%%%%%%%%%%%%%%%%%%%%%%%%%%%%%%%%%%%%%%
%%%%%%%%%%%%%%%%%%%%%%%%%%%%%%%%%%%%%%%%%%%%%%%%%%%%%%%%%%%%%%%%%%%%%%%%%%%%%%%%%%%%%%%%%%
%%%%%%%%%%%%%%%%%%%%%%%%%%%%%%%%%%%%%%%%%%%%%%%%%%%%%%%%%%%%%%%%%%%%%%%%%%%%%%%%%%%%%%%%%%

\section*{Acknowledgements}
S.C. acknowledges INFN Sez. di Napoli (Iniziative Specifiche CQSKY and TEONGRAV) for
financial support. O.L. wishes to thank Manuel Scinta for the help in the numerical
analysis. O.L. is financially supported by the European PONa3 00038F1 KM3NeT (INFN)
Project. The research of E.N.S. is implemented within the framework of the
Operational Program ``Education and Lifelong Learning'' (Actions
Beneficiary: General Secretariat for Research and Technology), and
is co-financed by the European Social Fund (ESF) and the Greek State.

\end{document}